    \documentclass[twocolumn,final]{svjour3}
\smartqed  
\usepackage{graphicx}
 \usepackage{mathptmx}      
    \usepackage{bm}

    \usepackage{dcolumn}

    \usepackage{amsmath}

    \usepackage{amssymb}

\newcommand\beq{\begin{equation}}
\newcommand\eeq{\end{equation}}
\newcommand\beqa{\begin{eqnarray}}
\newcommand\eeqa{\end{eqnarray}}
\newcommand{\nn}{\nonumber\\}

\newcommand{\dd}{{\text{d}}}

\newcommand{\Ia}{{\text{Ia}}}
\newcommand{\Ib}{{\text{Ib}}}
\newcommand{\Ih}{{\text{Ih}}}
\newcommand{\IIa}{{\text{IIa}}}
\newcommand{\IIb}{{\text{IIb}}}

\newcommand{\LL}[1]{\mathcal{L}_{a_2,a_3}\left\{#1\right\}}
\newcommand{\Ll}[1]{\mathcal{L}_{a_2}\left\{#1\right\}}

 \journalname{Granular Matter}
\begin{document}

\title{The second and third Sonine coefficients  of a freely cooling granular gas revisited
}

\titlerunning{The second and third Sonine coefficients}        

\author{Andr\'es Santos         \and
        Jos\'e Mar\'{\i}a Montanero
}


\institute{A. Santos \at
              Departamento de F\'{\i}sica, Universidad
de Extremadura, E-06071 Badajoz, Spain \\
              Tel.: +34-924289540\\
              Fax: +34-924289651\\
              \email{andres@unex.es}           
           \and
           Jos\'e Mar\'{\i}a Montanero \at
              Departamento de Ingenier\'{\i}a
Mec\'anica, Energ\'etica y de los Materiales, Universidad de
Extremadura, E-06071 Badajoz, Spain\\
              Tel.: +34-924289600\\
              Fax:  +34-924289601\\
              \email{jmm@unex.es}
}

\date{\today}

\maketitle

\begin{abstract}
In its simplest statistical-mechanical description, a granular fluid
can be modeled as composed of smooth inelastic hard spheres (with a
constant coefficient of normal restitution $\alpha$) whose velocity
distribution function obeys the Enskog-Boltzmann equation. The basic
state of a granular fluid is the homogeneous cooling state,
characterized by a homogeneous, isotropic, and stationary
distribution of scaled velocities, $F(\mathbf{c})$.  The behavior of $F(\mathbf{c})$ in the
domain of thermal velocities ($c\sim 1$) can be characterized by the
two first non-trivial coefficients ($a_2$ and $a_3$) of an expansion
in Sonine polynomials. The main goals of this paper are  to review
some of the previous efforts made to estimate (and measure
in computer simulations) the
$\alpha$-dependence of $a_2$ and $a_3$,  to report new computer simulations
results of $a_2$ and $a_3$ for two-dimensional systems, and  to investigate the possibility of
proposing theoretical estimates of $a_2$ and $a_3$ with an optimal
compromise between simplicity and accuracy.
\end{abstract}

\section{Introduction}
The prototype model for a granular gas is a system of smooth
inelastic hard spheres characterized by a coefficient of normal
restitution $0<\alpha\leq 1$ \cite{C90,BP04}. Kinetic theory arguments similar to
those followed in the elastic case allow one to derive the Boltzmann
and Enskog equations \cite{BDS97} for the velocity distribution
function $f(\mathbf{r},\mathbf{v},t)$.

Perhaps the basic and simplest physical state for a granular gas is
the homogeneous cooling state (HCS), where the gas is isolated and
has an isotropic and  spatially uniform distribution \cite{BP04}. In this state,
the collisional loss of energy, characterized by the cooling rate $\zeta$, makes the mean kinetic energy
(directly related to the so-called granular temperature $T$) decay
monotonically in time following Haff's law \cite{H83}:
\beq
T(t)=\frac{T(0)}{\left[1+\frac{1}{2}\zeta(0)t\right]^2}.
\label{Haff}
\eeq
 Therefore,
the distribution function evolves in time toward a delta function, i.e.,
$f(\mathbf{v})\to n \delta(\mathbf{v})$, where $n$ is the number
density. However, the simplicity of this trivial asymptotic state is
deceiving since the distribution function actually reaches an
interesting scaling (or self-similar) form
\beq
f(\mathbf{v},t)=n v_0^{-d}(t)F(\mathbf{c}(t)),\quad
\mathbf{c}(t)=\mathbf{v}/v_0(t).
\label{1}
\eeq
Here, $d$ is the dimensionality and $v_0(t)$ is the thermal speed
defined by
\beq
\frac{d}{2}v_0^{2}(t)=\frac{1}{n}\int \dd\mathbf{v}\, v^2
f(\mathbf{v},t).
\label{2}
\eeq
By definition,
\beq
\langle c^2\rangle=\frac{d}{2},
\label{6}
\eeq
where the (scaled) velocity moments are
\beq
\langle c^{2p}\rangle=\int \dd\mathbf{c}\, c^{2p} F(\mathbf{c}).
\label{7}
\eeq

In the HCS, the Enskog--Boltzmann equation for the probability
distribution function $F(\mathbf{c})$ of the reduced velocity is
\cite{vNE98}
\beq
\frac{\mu_2}{d}
\frac{\partial}{\partial \mathbf{c}}\cdot \mathbf{c}
F(\mathbf{c})=I[\mathbf{c}|F,F],
\label{3}
\eeq
where the collision operator is
\begin{eqnarray}
\label{4}
{I}[\mathbf{c}_1|{F},{F}]&=&\int \dd\mathbf{c}_2\int
\dd\widehat{\bm{\sigma}}\,
\Theta(\mathbf{c}_{12}\cdot\widehat{\boldsymbol{\sigma}})
(\mathbf{c}_{12}\cdot\widehat{\bm{\sigma}})\nonumber\\
&&\times\left[\alpha^{-2}{F}(\mathbf{c}_1''){F}(\mathbf{c}_2'')-
{F}(\mathbf{c}_1){F}(\mathbf{c}_2)\right]
\end{eqnarray}
and we have introduced the collisional moments
\begin{equation}
\label{5}
\mu_{2p}\equiv - \int \dd\mathbf{c}\,c^{2p}{I}[\mathbf{c}|{F},{F}],
\end{equation}
so that $2\mu_2/d$  is the dimensionless cooling rate. In Eq.\
(\ref{4}), $\mathbf{c}_{12}\equiv \mathbf{c}_1-\mathbf{c}_2$ is the
relative velocity of the colliding particles,
$\widehat{\bm{\sigma}}$ is a unit vector directed along the
line of centers from the sphere $1$ to the sphere $2$, $\Theta$ is
the Heaviside step function, and $(\mathbf{c}_1'',\mathbf{c}_2'')$
are the precollisional or restituting velocities yielding
$(\mathbf{c}_1,\mathbf{c}_2)$ as the postcollisional ones, i.e.
\beq
\mathbf{c}_{1,2}''=\mathbf{c}_{1,2}\mp\frac{1}{2}(1+\alpha^{-1})
(\mathbf{c}_{12}\cdot\widehat{\bm{\sigma}})
\widehat{\bm{\sigma}}.
\label{8}
\eeq

The exact solution of Eq.\ (\ref{3}) is not known. Except in the
elastic case ($\alpha=1$), the Maxwellian
$F(\mathbf{c})=\pi^{-d/2}e^{-c^2}\equiv \phi(\mathbf{c})$ is not a
solution. In particular, if $\alpha<1$, it is known that
$F(\mathbf{c})$ develops an exponential high-energy tail
\cite{vNE98,EP97},
\beq
F(\mathbf{c})\sim e^{-\xi c},\quad \xi=\frac{d\pi^{(d-1)/2}}{\Gamma(\frac{d+1}{2})\mu_2}.
\label{tail}
\eeq
 A convenient way of
characterizing the deviation of $F(\mathbf{c})$ from
$\phi(\mathbf{c})$ in the regime of low and intermediate speeds is
through the Sonine polynomial expansion
\beq
F(\mathbf{c})=\phi(\mathbf{c})\left[1+\sum_{k=2}^\infty a_k
L_k^{(\frac{d-2}{2})}(c^2) \right],
\label{9}
\eeq
where $L_k^{(a)}$ are generalized Laguerre (or Sonine) polynomials
\cite{AS72}. The  first two non-trivial coefficients are $a_2$ and
$a_3$. They are related to the fourth and sixth velocity moments as
\begin{equation}
\label{10}
\langle c^4\rangle=\frac{d(d+2)}{4}\left(1+a_2\right),
\end{equation}
\beq
\label{11}
\langle c^6\rangle=\frac{d(d+2)(d+4)}{8}\left(1+3a_2-a_3\right).
\eeq

{The Sonine expansion \eqref{9} is known to be only asymptotic \cite{NBSG07}, so that its practical applicability  is restricted to low and intermediate velocities (say $c\lesssim 1$), in which case the most relevant coefficients are $a_2$ and $a_3$.  Therefore, the determination of these two coefficients is important to quantify the basic deviations of the HCS distribution $F(\mathbf{c})$ from the Maxwellian $\phi(\mathbf{c})$, at least for $c\lesssim 1$. This explains the interest  this problem has attracted over the years \cite{BP04,vNE98,NBSG07,GS95,BRMC96,GD99,MS00,BP00,HOB00,MG02,CDPT03,BP06,AP06}. Apart from this formal motivation, the knowledge of $a_2(\alpha)$ and $a_3(\alpha)$, especially in the case of the former, is needed to evaluate the dependence of the transport coefficients on inelasticity \cite{NBSG07,BDKS98,GSM07}.}

The aim of this paper is three-fold. First, some of the efforts done in the last dozen years or so to estimate $a_2$ and $a_3$ by theoretical tools and to measure them in simulations are briefly reviewed in Sec.\ \ref{sec2}. Next, we explore the possibility of deriving theoretical expressions for $a_2$ and $a_3$ with an optimal compromise between simplicity and accuracy. To that end, we restrict ourselves to the class of linear approximations, revisit some of the ones already proposed in the literature, and construct a few new ones in Sec.\ \ref{sec3}. Those approximations are compared with new ($d=2$) and recently published \cite{BP06} ($d=3$) computer simulations  in Sec.\ \ref{sec4}. Section \ref{sec5} addresses the case of granular gases heated with a white-noise thermostat. Finally, the conclusions are summarized in Sec.\ \ref{sec6}, while some complementary material is relegated to the Appendices.

\section{A brief  review of previous results}
\label{sec2}

{Taking (even) moments in both sides
of Eq.\ (\ref{3}) one gets the \emph{exact} set of moment equations}
\begin{equation}
\label{12}
\mu_{2p}=p\mu_2 \frac{\langle c^{2p}\rangle}{\langle c^2\rangle}, \quad p\geq 2,
\end{equation}
{where use has been made of Eq.\ \eqref{6}.
The condition $p\geq 2$ is introduced because Eq.\ (\ref{12})
becomes an identity for $p=0$ and also for $p=1$.}

{It is important to notice that the collisional moments are functionals of the distribution function, so that Eq.\ \eqref{12} implies a coupling among \emph{all} the Sonine coefficients $a_k$ and there is no \emph{a priori} reason to expect a chosen truncation to provide accurate results for a subset of coefficients. On the other hand, most of the routes followed to get theoretically based results assume some kind of truncation and/or order-of-magnitude simplification. More specifically, one
usually \emph{approximates} the first few collisional moments $\mu_{2p}$  by
inserting the expansion (\ref{9}) into Eqs.\ (\ref{4}) and
(\ref{5}), truncating the expansion after a certain order and, in
some cases, neglecting nonlinear terms. The resulting set of
approximate equations is then solved algebraically to obtain
estimates for the desired coefficients $a_k$.
In principle, these estimates are uncontrolled and can be assessed only after comparison with computer simulations.}

The first application of this method was carried out by Goldshtein and Shapiro in a pioneering and extensive paper \cite{GS95}. They derived a simple expression for $a_2$ in the three-dimensional ($d=3$) case by taking a \emph{linear} approximation (namely, neglecting $a_2^2$ and $a_k$ with $k\geq 3$) in Eq.\ \eqref{12} with $p=2$. Here we quote their result:
\beq
a_2=\frac{16(1-\alpha)(1-2\alpha^2)}{\lambda_0+\lambda_1\alpha+\lambda_2\alpha^2(1-\alpha)},
\label{GS}
\eeq
where $(\lambda_0,\lambda_1,\lambda_2)=(401,-337,190)$. They noted that, according to their estimate, the magnitude of $a_2$ was quite small ($|a_2|<0.04$). However, there was a small algebraic mistake in their derivation that was subsequently corrected by van Noije and Ernst \cite{vNE98}, who also generalized the result to arbitrary $d$. The expression derived by van Noije and Ernst (vNE) maintains the form \eqref{GS}, except that $(\lambda_0,\lambda_1,\lambda_2)=(9+24d,8d-41,30)$, what in the three-dimensional case becomes $(\lambda_0,\lambda_1,\lambda_2)=(81,-17,30)$. This yields values of $|a_2|$ up to five times larger than those predicted by the (wrong) original expression by Goldshtein and Shapiro. Although published in 1998, vNE's result had been circulating earlier and so in 1996 Brey et al.\ \cite{BRMC96} confirmed  its accuracy for $d=3$ and $\alpha\geq 0.7$ by comparison with DSMC simulations \cite{DSMC} of the Boltzmann equation. Brey et al.\ \cite{BRMC96} also presented simulation data for $\langle c^6\rangle$ but they did not extract from them the corresponding values of $a_3$. When this is done from Figs.\ 5 and 6 of Ref.\ \cite{BRMC96}, one gets  $a_3\simeq -0.005$ for $0.7\leq \alpha\leq 0.9$ and $d=3$.
{More recently, Ahmad and Puri \cite{AP06} carried out large-scale event-driven molecular dynamics simulations and measured $a_2$ and $a_3$ in the HCS for $\alpha\geq 0.7$ in two- and three-dimensional systems. The results confirmed the accuracy of vNE's expression of $a_2$ in that range of inelasticity and showed that $a_3$ was typically four to five times smaller than $a_2$. These authors also studied the time evolution of $a_k$ for $k=2$--$5$ to monitor the transition from the HCS to an inhomogeneous cooling state.}

In 1999, Garz\'o and Dufty \cite{GD99} studied the HCS for three-dimensional binary mixtures. Neglecting again $a_2^2$ and  $a_k$ with $k\geq 3$,  they obtained explicit expressions for the Sonine coefficient $a_2$ of both species as functions of the three coefficients of restitution and of the temperature ratio $T_1/T_2$. To close the problem, it is necessary to determine $T_1/T_2$ from the condition of equal cooling rates for both species, yielding a highly nonlinear equation. The results showed that typically the component made of particles with a larger mass has a higher temperature and a higher value of $a_2$. The theoretical predictions were later validated by  DSMC simulations \cite{MG02}.

The authors pointed out in 2000 \cite{MS00} that the linear approximation to get $a_2$ is not univocally defined, the result depending on the way that the quantities in Eq.\ \eqref{12} are arranged. In particular, if Eq.\ \eqref{12} for $p=2$ is rewritten as $\mu_4/\langle c^4\rangle=2\mu_2/\langle c^2\rangle$ and then a linear approximation is applied the result is given again by Eq.\ \eqref{GS} but with $(\lambda_0,\lambda_1,\lambda_2)=(25+24d,8d-57,-2)$, implying $(\lambda_0,\lambda_1,\lambda_2)=(97,-33,-2)$ for $d=3$. This alternative expression is hardly distinguishable from vNE's if $\alpha\gtrsim 0.5$ but provides up to 16\% smaller values than the latter for higher inelasticities. We performed DSMC simulations of $a_2$ for $d=3$ and $\alpha\geq 0.2$ which exhibited an excellent agreement with our alternative linear approximation. Moreover, DSMC data of $a_3$ were also presented in Ref.\ \cite{MS00}. While $a_3\simeq -0.005$ for $0.6<\alpha<0.9$,  a relatively rapid decay of $a_3$ for higher inelasticities was observed.

Brilliantov and P\"oschel \cite{BP00} were possibly the first ones to depart from the linear approximation. They neglected $a_k$ with $k\geq 3$ but retained $a_2^2$ in Eq.\ \eqref{12} with $p=2$, thus obtaining a closed cubic equation for $a_2$ (and $d=3$). Its physical root is very close to vNE's expression for $\alpha\gtrsim 0.4$ but becomes up to 10\% larger than it  for smaller values of $\alpha$.  Taking into account the simulation results presented in Ref.\ \cite{MS00}, it turns out that the physical root of the cubic equation deviates in the wrong direction from vNE's simpler linear approximation. This paradoxical outcome shows that the Sonine coefficients $a_k$ with $k\geq 3$ are not negligible for $\alpha\lesssim 0.4$.

A different truncation scheme was followed by Huthmann et al.\ \cite{HOB00}, who assumed that $a_k=\mathcal{O}(\epsilon^k)$, where $\epsilon\sim |a_2|^{1/2}$ was treated as a small parameter. The solution to order $k\geq 2$ was obtained by taking Eq.\ \eqref{12} for $p=2,\ldots, k$ and formally neglecting terms of order $\epsilon^\ell$ with $\ell>k$.  The second-order solution recovers the vNE result for $a_2$. In the third-order solution one has a set of two linear equations for $a_2$ and $a_3$, but  the fourth-order approximation involves a set of three equations for $a_2$, $a_3$, and $a_4$ that include $a_2^2$, so that the problem becomes nonlinear for $k\geq 4$. In this approach, the coefficient $a_2$ is renormalized as the truncation order increases. Huthmann et al.\ applied their scheme to $d=2$ and observed that the values of $a_2$ obtained from order $\epsilon^2$ to order $\epsilon^6$ were practically the same as long as $\alpha\gtrsim 0.6$. However, those values were dramatically sensitive to the truncation order for higher inelasticities, thus indicating that the assumption $a_k=\mathcal{O}(\epsilon^k)$ fails if $\alpha\lesssim 0.6$. Molecular dynamics simulations showed a good performance of vNE's expression for hard disks and $\alpha\geq 0.4$.

Coppex et al.\ \cite{CDPT03} tried an approach to estimate $a_2$ different from those based on Eq.\ \eqref{12}. They started from Eq.\ \eqref{3} in the limit $\mathbf{c}\to \mathbf{0}$ and  then inserted the Sonine expansion \eqref{9} by neglecting $a_2^2$ and $a_k$ with $k\geq 3$. The solution of the resulting linear equation for $a_2$ had the structure of a polynomial of fourth degree in $\alpha^2$ divided by a polynomial of eighth degree in $\alpha$ (with no $\alpha^5$ and $\alpha^7$ terms). Although promising, this alternative method yields poor results for small and moderate inelasticities and only improves over the vNE benchmark formula if $\alpha\lesssim 0.4$, as comparison with DSMC data for $d=2$ shows \cite{CDPT03}. Coppex et al.\ also elaborated further on the ambiguity of the linear approximation for $a_2$ pointed out in Ref.\ \cite{MS00}. In particular, they showed that the linear approximation as applied to $\mu_4/\langle c^4\rangle=2\mu_2/\langle c^2\rangle$ and to $\mu_4\langle c^2\rangle/2\mu_2\langle c^4\rangle=1$ presented a very good agreement with their two-dimensional simulations.

More recently, Brilliantov and P\"oschel \cite{BP06} have considered the linear approximation of Eq.\ \eqref{12} with $p=2$ and $p=3$ by neglecting $a_2^2$, $a_2a_3$, $a_3^2$, and $a_k$ with $k\geq 4$. This gives a set of two linear equations for $a_2$ and $a_3$ that is actually equivalent to Huthmann et al.'s method to order $\epsilon^3$. By comparing with their own DSMC simulations for $d=3$, Brilliantov and P\"oschel showed that their expression of $a_2$, while rather more complicated than vNE's, provided worse estimates than the latter for $\alpha\lesssim 0.6$. As for  their expression of $a_3$, it was quite good for $\alpha\gtrsim 0.6$ and exhibited the right qualitative behavior for larger inelasticities. Apart from $a_2$ and $a_3$, they also measured $a_4$, $a_5$, and $a_6$ in the simulations, showing that their values were not negligible if $\alpha\lesssim 0.6$. In fact,  these authors argued that the Sonine expansion breaks down for large inelasticity.

Using the asymptotic high-velocity tail \eqref{tail}, Noskowicz et al.\ \cite{NBSG07} have shown that $a_k\propto (-4/\xi^2)^k  (k+1)!$ for large $k$, so that the series \eqref{9} is divergent, although it is asymptotic and Borel resummable. On the other hand, the Sonine expansion of the modified function $G_\gamma(\mathbf{c})=e^{-(1-\gamma)c^2}F(\mathbf{c})$ does converge for $0<\gamma<\frac{1}{2}$. Truncating the Sonine expansion of $G_\gamma(\mathbf{c})$ after $k=N_s$ (with $N_s=10$, $20$, and $40$), Noskowicz et al.\ obtained numerically the associated  Sonine coefficients $a_k^{(\gamma)}$, $k=0,1,\ldots N_s$, with the help of symbolic software. Once $G_\gamma(\mathbf{c})$ is (approximately) determined in this way, the Sonine coefficients $a_k$ of the true distribution function $F(\mathbf{c})=e^{(1-\gamma)c^2}G_\gamma(\mathbf{c})$ can be obtained by quadratures. The numerical results for $a_2$, which are well fitted in the three-dimensional case by Eq.\ \eqref{GS} with $(\lambda_0,\lambda_1,\lambda_2)=(104.1,-51.43,78.67)$, confirmed that vNE's expression overestimates $a_2$ for $\alpha\lesssim 0.5$, while the alternative expression proposed in Ref.\ \cite{MS00}  is rather accurate (although it slightly overestimates $a_2$).

\section{Theoretical estimates from linear approximations\label{sec3}}
Our main goal now is to get \emph{estimates} of  the
Sonine coefficients $a_2$ and $a_3$ by the application of
approximations that neglect the coefficients $a_k$ with $k\geq 4$ as
well as the nonlinear terms  $a_2^2$, $a_2a_3$, and $a_3^2$. As we
will see, this recipe is not a systematic method and so it does not
provide a unique result.

Given a functional $X[F]$ of the scaled probability distribution
function $F(\mathbf{c}$), henceforth we will use the notation
$\LL{X}$ to denote an approximation to $X[F]$ obtained by using Eq.\
\eqref{9} and neglecting  $a_k$ with $k\geq 4$ and nonlinear terms
(like $a_2^2$, $a_2a_3$, and $a_3^2$). Furthermore, if $a_3$ is also
neglected, the corresponding approximation will be denoted by
$\Ll{X}$. In particular, in the case of the collisional moments
defined by Eq.\ \eqref{5} with $p=1$, 2, and 3, one
gets
\beq
\LL{\mu_2}=
A_0+A_2a_2+A_3a_3,
\label{15}
\eeq
\beq
\LL{\mu_4}=
B_0+B_2a_2+B_3a_3,
\label{16}
\eeq
\beq
\LL{\mu_6}=
C_0+C_2a_2+C_3a_3.
\label{17}
\eeq
The expressions for the coefficients $A_i$, $B_i$, and $C_i$ as
functions of $\alpha$ and $d$ were derived  by van Noije and Ernst
\cite{vNE98} and by Brilliantov and P\"oschel \cite{BP04,BP06}. They
are given in  Appendix \ref{appA}.
Obviously, $\Ll{\mu_2}$ and  $\Ll{\mu_4}$ are obtained by formally setting $A_3\to 0$ and $B_3\to 0$ on the right-hand sides of Eqs.\ \eqref{15} and \eqref{16}, respectively.

The exact equation \eqref{12} becomes an approximation when it is
linearized with respect to $a_2$ and $a_3$. For $p=2$ and $p=3$ we get
\beqa
0&=&\LL{\mu_4-2\mu_2\frac{\langle c^4\rangle}{\langle c^2\rangle}}\nn
&=&B_0-(d+2)A_0+\left[B_2-(d+2)(A_0+A_2)\right]a_2\nn
&&+[B_3-(d+2)A_3]a_3,
\label{n1}
\eeqa
\beqa
0&=&\LL{\mu_6-3\mu_2\frac{\langle c^6\rangle}{\langle c^2\rangle}}\nn
&=&C_0-\frac{3}{4}(d+2)(d+4)A_0+[C_2-\frac{3}{4}(d+2)(d+4)\nn
&&\times(3A_0+A_2)]a_2+[C_3-\frac{3}{4}(d+2)(d+4)(A_3-A_0)]a_3.\nn
&&
\label{n2}
\eeqa

The non-systematic character of the linearization method is made
evident if one proceeds in the same way, except that Eq.\ \eqref{12}
is rewritten in the equivalent form
\begin{equation}
\label{12bis}
\frac{\mu_{2p}}{\langle c^{2p}\rangle}=p\frac{\mu_2}{\langle c^{2}\rangle} , \quad p\geq 2.
\end{equation}
This equation shows that the \emph{rescaled} collisional moment ${\mu_{2p}}/{\langle c^{2p}\rangle}$ is just proportional to $p$.
Now, instead of Eqs.\ \eqref{n1} and \eqref{n2} we have
\beqa
0&=&\LL{\frac{\mu_4}{\langle
c^4\rangle}-2\frac{\mu_2}{\langle c^{2}\rangle}}\nn
&=&B_0-(d+2)A_0+[B_2-B_0-(d+2)A_2]a_2\nn
&&+[B_3-(d+2)A_3]a_3,
\label{n1b}
\eeqa
\beqa
0&=&\LL{\frac{\mu_6}{\langle
c^6\rangle}-3\frac{\mu_2}{\langle c^{2}\rangle}}\nn
&=&C_0-\frac{3}{4}(d+2)(d+4)A_0+\Big[C_2-3C_0-\frac{3}{4}(d+2)\nn
&&\times (d+4) A_2\Big]a_2+
\Big[C_3+C_0-\frac{3}{4}(d+2)(d+4)A_3\Big]a_3.\nn
&&
\label{n2b}
\eeqa
Note that Eq.\ \eqref{n1b} is  obtained from Eq.\ \eqref{n1} if one
formally replaces $(d+2)A_0$ by $B_0$ in the coefficient of $a_2$.
Likewise, Eq.\ \eqref{n2b} is  obtained from Eq.\ \eqref{n2} by
formally replacing  $\frac{3}{4}(d+2)(d+4)A_0$ by $C_0$ in the
coefficients of $a_2$ and $a_3$. Nevertheless, the approximations
\eqref{n1} and \eqref{n2} are  different from the approximations \eqref{n1b} and \eqref{n2b}, respectively, so they provide different estimates of the coefficients
$a_2$ and $a_3$. Henceforth we will refer to the approximations
\eqref{n1} and \eqref{n2}, which are based on Eq.\ \eqref{12}, with
the label ``a''. Analogously, the approximations \eqref{n1b} and
\eqref{n2b}, which are based on Eq.\ \eqref{12bis}, will be  labeled
by ``b''. Of course, other alternative ways of rewriting Eq.\ \eqref{12} are possible \cite{MS00,CDPT03}. With independence of the
adopted approach (say ``a'' or ``b''), there are two basic classes of approximations: either  $a_3$ is neglected
versus $a_2$  in the equation for $\mu_4$ (but not in the equation for $\mu_6$) or both Sonine coefficients are treated on the same footing.

\subsection{Class-I approximations: $a_3\ll a_2$ }
Let us first assume that $a_3$ can be neglected versus $a_2$ in
either Eq.\ \eqref{n1} (approach ``a'') or Eq.\ \eqref{n1b} (approach
``b''):
\beqa
0&=&\Ll{\mu_4-2\mu_2\frac{\langle c^4\rangle}{\langle c^2\rangle}}\nn
&=&B_0-(d+2)A_0+\left[B_2-(d+2)(A_0+A_2)\right]a_2,
\label{n1I}
\eeqa
\beqa
0&=&\Ll{\frac{\mu_4}{\langle
c^4\rangle}-2\frac{\mu_2}{\langle c^{2}\rangle}}\nn
&=&B_0-(d+2)A_0+[B_2-B_0-(d+2)A_2]a_2.
\label{n1bI}
\eeqa
These are linear
equations for $a_2$ whose respective solutions are
\beqa
a_2^\Ia(\alpha)&=&\frac{B_0-(d+2)A_0}{(d+2)(A_2+A_0)-B_2}\nn &=&
\frac{16(1-\alpha)(1-2\alpha^2)}{9+24d-(41-8
d)\alpha+30(1-\alpha)\alpha^2},
\label{23}
\eeqa
\beqa
a_2^\Ib(\alpha)&=&\frac{B_0-(d+2)A_0}{(d+2)A_2-(B_2-B_0)}\nn &=&
\frac{16(1-\alpha)(1-2\alpha^2)}{25+24d- (57-
8d)\alpha-2(1-\alpha)\alpha^2},
\label{24}
\eeqa
where in the last steps use has been made of the explicit
expressions of $A_0$, $A_2$, $B_0$, and $B_2$. As recalled in Sec.\ \ref{sec2}, the method labeled
here as ``Ia'' was the one first followed by Goldshtein and Shapiro
\cite{GS95}, the corresponding estimate, Eq.\  \eqref{23}, being
first obtained by van Noije and Ernst \cite{vNE98}. The alternative
method ``Ib'', Eq.\ \eqref{24}, was  proposed in Ref.\ \cite{MS00}. It
is interesting to note that
\beq
a_2^\Ib=\frac{a_2^\Ia}{1+a_2^\Ia}.
\label{n7}
\eeq
Comparison with DSMC data shows that the estimate $a_2^\Ib$ is
superior to $a_2^\Ia$ for $\alpha \lesssim 0.6$ \cite{MS00}.
Other class-I approximations for $a_2$   were considered by Coppex et al.\ \cite{CDPT03} and are more generally described in Appendix \ref{appB}.

Once $a_3$ has been neglected in Eqs.\ \eqref{n1} and \eqref{n1b}, we
can use Eq.\ \eqref{n2} (approach ``a'') or Eq.\ \eqref{n2b} (approach ``b'')
to express $a_3$ in terms of $a_2$. The respective results are
\beq
a_3^\Ia(\alpha)=G_a(\alpha,a_2^\Ia(\alpha)),
\label{n8}
\eeq
\beq
a_3^\Ib(\alpha)=G_b(\alpha,a_2^\Ib(\alpha)),
\label{n9}
\eeq
where
\beqa
G_a(\alpha,a_2)&\equiv&
\Big\{C_0-\frac{3}{4}(d+2)(d+4)A_0+\Big[C_2-\frac{3}{4}(d+2)\nn
&&\times(d+4)(3A_0+A_2)\Big]a_2\Big\}/\Big[\frac{3}{4}(d+2)(d+4)\nn
&&\times(A_3-A_0)-C_3\Big],
\label{n10}
\eeqa
\beqa
G_b(\alpha,a_2)&\equiv&
\Big\{C_0-\frac{3}{4}(d+2)(d+4)A_0+\Big[C_2-3C_0\nn
&&-\frac{3}{4}(d+2)
(d+4)A_2\Big]a_2\Big\}/\Big[\frac{3}{4}(d+2)\nn
&&\times(d+4)A_3-C_3-C_0\Big].
\label{n11}
\eeqa

It is also possible to construct a \emph{hybrid} approximation
``Ih'' in which $a_2$ is obtained from Eq.\ \eqref{n1bI} and $a_3$ is subsequently obtained from Eq.\ \eqref{n2}.  In that case, $a_2^\Ih=a_2^\Ib$ while
\beq
a_3^\Ih(\alpha)=G_a(\alpha,a_2^\Ib(\alpha)).
\label{n12}
\eeq
The other hybrid possibility $a_3=G_b(\alpha,a_2^\Ia(\alpha))$ turns
out to be rather poor and will not be further considered here.

\subsection{Class-II approximations: $a_3\sim a_2$ }
If $a_3$ is formally treated as being of the same order as $a_2$,
Eqs.\ \eqref{n1} and \eqref{n2} become a linear set
of two coupled equations for $a_2$ and $a_3$ (approach ``a''). This was
the method recently considered by Brilliantov and P\"oschel
\cite{BP06}. Now the problem is algebraically more involved than in
the class-I approximation. The solution for $a_2$ is
\beq
a_2^\IIa(\alpha)=\frac{{N}_a(\alpha)}{{D}_a(\alpha)},
\label{26}
\eeq
where
\beqa
{N}_a(\alpha)&\equiv& {C_3}{B_0}- {C_0}{B_3} +
    (d+2)( A_3 C_0- A_0 C_3)  +
       \frac{3}{4}(d+2)\nn
       &&\times
       (d+4)[A_0 B_3-  ({A_3}-A_0){B_0}-(d+2)A_0^2],
       \label{28}
\eeqa
\beqa
     {D}_a(\alpha)&\equiv& C_2 B_3-C_3 B_2 +
   (d+2)[ ({A_2}+A_0){C_3}-A_3 C_2]\nn
   &&+\frac{3}{4}(d+2)(d+4)
       [(A_3-A_0)B_2-(A_2+3A_0)B_3\nn
       &&+(d+2)(A_0+A_2+2A_3)A_0].
       \label{29}
\eeqa
The corresponding result for $a_3$ is
\beq
a_3^\IIa(\alpha)=G_a(\alpha, a_2^\IIa(\alpha)).
\label{27}
\eeq
Note that, although the same functional form $G_a(\alpha,a_2)$
appears in Eqs.\ \eqref{n8} and \eqref{27}, obviously $a_3^\IIa(\alpha)\neq
a_3^\Ia(\alpha)$.

The same class-II approximation can be applied to Eqs.\ \eqref{n1b}
and \eqref{n2b} (approach ``b''). The solution is now
\beq
a_2^\IIb(\alpha)=\frac{{N}_b(\alpha)}{{D}_b(\alpha)},
\label{30}
\eeq
\beq
a_3^\IIb(\alpha)=G_b(\alpha, a_2^\IIb(\alpha)),
\label{33}
\eeq
where
\beqa
{N}_b(\alpha)&\equiv& {(C_0+C_3)}{B_0}- {C_0}{B_3} +
    (d+2)[ A_3C_0-(C_0+C_3)\nn
    && \times A_0]+
       \frac{3}{4}(d+2)(d+4)(A_0 B_3-  {A_3}{B_0}),
       \label{31}
\eeqa
\beqa
     {D}_b(\alpha)&\equiv& (C_2-3C_0) B_3-(C_0+C_3) (B_2-B_0) +
   (d+2)\nn
   &&\times[ {A_2}(C_0+C_3)-A_3 (C_2-3C_0)]\nn
       &&+\frac{3}{4}(d+2)(d+4)\left[A_3(B_2-B_0)-A_2B_3\right].
       \label{32}
\eeqa

\begin{table*}
\caption{Summary of the linear approximations considered in this paper}
\label{table1}       
\begin{tabular}{cccccccccc}
\hline\noalign{\smallskip}
 & &\multicolumn{2}{c}{References}&&\multicolumn{2}{c}{Behavior of $a_2$}&&\multicolumn{2}{c}{Behavior of $a_3$}\\
\cline{3-4}\cline{6-7}\cline{9-10}
Label &Equations&$a_2$&$a_3$&&$0<\alpha<0.6$&$0.6<\alpha<1$&&$0<\alpha<0.6$&$0.6<\alpha<1$ \\
\noalign{\smallskip}\hline\noalign{\smallskip}
Ia&$\begin{array}{l}\Ll{\mu_4-2\mu_2\langle c^4\rangle/\langle c^2\rangle}=0\\
\LL{\mu_6-3\mu_2\langle c^6\rangle/\langle c^2\rangle}=0$ $\end{array}$&  \protect\cite{vNE98}&New&&Fair&Good&&Good&Fair\\
\\
 IIa&$\begin{array}{l}\LL{\mu_4-2\mu_2\langle c^4\rangle/\langle c^2\rangle}=0\\
\LL{\mu_6-3\mu_2\langle c^6\rangle/\langle c^2\rangle}=0$ $\end{array}$& \protect\cite{BP06}&\protect\cite{BP06}&&Fair&Good&&Fair&Good\\
\\
Ib&$\begin{array}{l}\Ll{\mu_4/\langle
c^4\rangle-2\mu_2/\langle c^2\rangle}=0\\
\LL{{\mu_6}/{\langle c^6\rangle}-3\mu_2/\langle c^2\rangle}=0$ $\end{array}$& \protect\cite{MS00}&New&&Good&Good&&Poor&Poor\\
\\
IIb&$\begin{array}{l}\LL{\mu_4/\langle
c^4\rangle-2\mu_2/\langle c^2\rangle}=0\\
\LL{\mu_6/\langle c^6\rangle-3\mu_2/\langle c^2\rangle}=0$ $\end{array}$& New&New&&Good&Good&&Poor&Good\\
\\
Ih&$\begin{array}{l}\Ll{\mu_4/\langle
c^4\rangle-2\mu_2/\langle c^2\rangle}=0\\
\LL{\mu_6-3\mu_2\langle c^6\rangle/\langle c^2\rangle}=0$ $\end{array}$& \protect\cite{MS00}&New&&Good&Good&&Good&Fair\\
\noalign{\smallskip}\hline
\end{tabular}
\end{table*}

The three class-I and two class-II approximations described in this Section are summarized in Table \ref{table1}. As said before, $a_2^\Ia$
and $a_2^\Ib=a_2^\Ih$ were already proposed in Refs.\ \cite{vNE98} and \cite{MS00}, respectively, while $a_2^\IIa$ and $a_3^\IIa$ were derived in Ref.\ \cite{BP06}. All the other possibilities, to the best of our knowledge, have not been considered before.

\section{Comparison with computer simulations\label{sec4}}
\begin{figure}
  \includegraphics[width=\columnwidth]{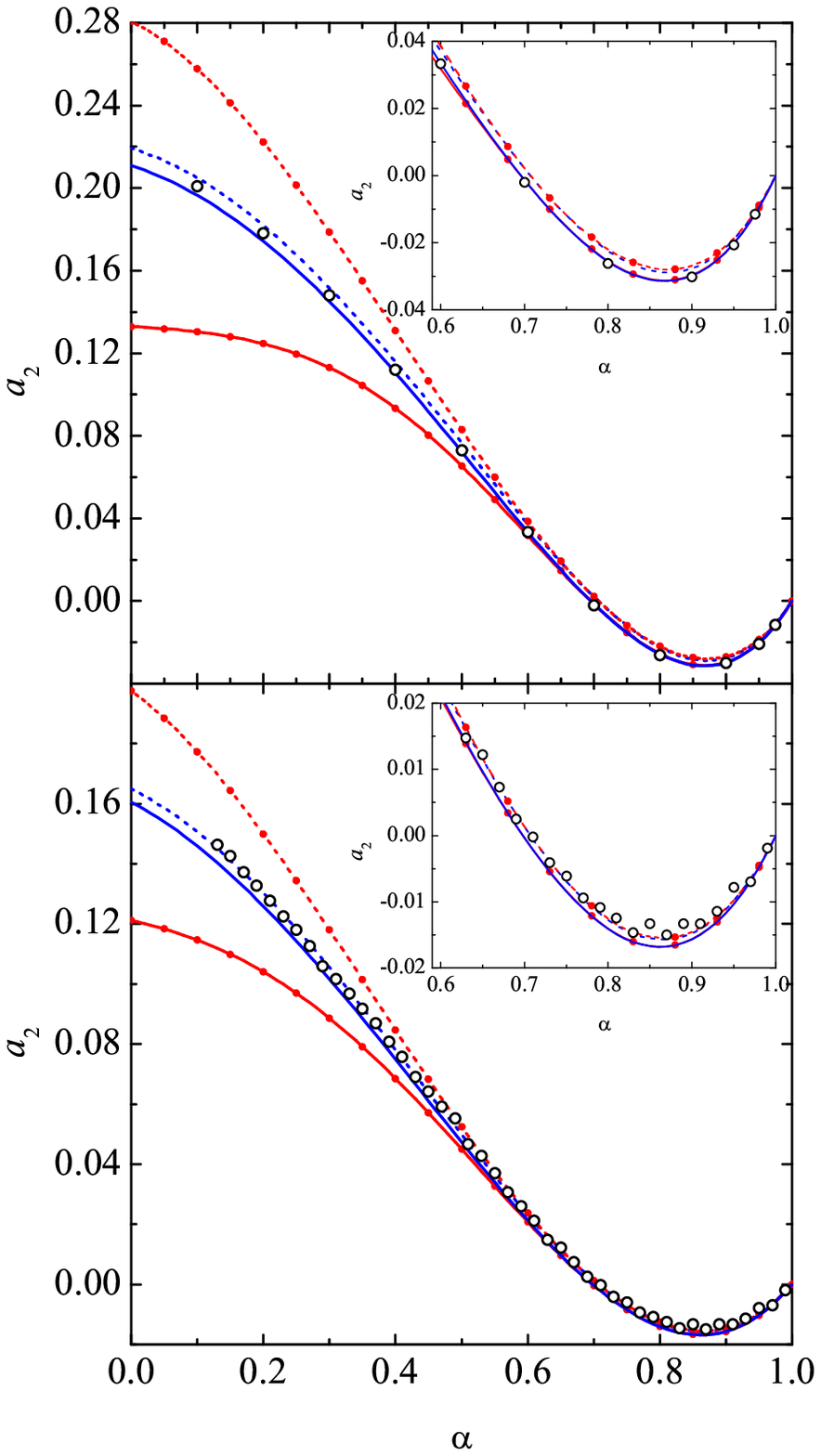}
\caption{(Color online)
Plot of the second Sonine coefficient $a_2$ as a function of the coefficient of restitution $\alpha$ for $d=2$ (top panel) and $d=3$ (bottom panel).
The circles represent DSMC results ($d=2$: this work; $d=3$:
Ref.\ \protect\cite{BP06}), while the lines correspond to $a_2^\Ia$
 (- -$\bullet$- -),
$a_2^\Ib=a_2^\Ih$ (- - -),
$a_2^\IIa$
(---$\bullet$---),
and $a_2^\IIb$
(------). The insets magnify the region $0.6\leq\alpha\leq 1$
}
\label{fig1}       
\end{figure}

In order to assess the reliability of the linear estimates for the Sonine coefficients $a_2$ and $a_3$ introduced in Sec.\ \ref{sec3}, it is necessary to compare them against computer simulations. To that end, we have performed new DSMC simulations for inelastic hard disks ($d=2$). In the case of inelastic hard spheres ($d=3$) we have used the extensive DSMC simulations presented in Ref.\ \cite{BP06}.
Figure \ref{fig1} compares the simulation data of $a_2$ with the theoretical estimates $a_2^\Ia$, $a_2^\Ib=a_2^\Ih$, $a_2^\IIa$, and $a_2^\IIb$. It is observed that the best global agreement with computer simulations is provided by the two approaches ``b'', i.e., the ones based on linearization of Eq.\ \eqref{12bis}, in contrast to the two approaches ``a'', which are based on linearization of Eq.\ \eqref{12}. Given that the class-I estimate $a_2^\Ib=a_2^\Ih$ (where $a_3$ is neglected) is much simpler than the class-II estimate $a_2^\IIb$ (where $a_3$ is retained), the former is preferable to the latter. In the region $0.6\leq \alpha\leq 1$ the four approximations practically coincide among themselves and with the simulation results, thus showing that $a_2^2$ and $a_k$ with $k\geq 3$ are indeed negligible in that region. On the other hand, our simulation data for hard disks ($d=2$) show a slight improvement of the two class-II approximations with respect to the two class-I approximations, what indicates  that the influence of $a_2^2$ is even smaller than that of $a_3$ for $0.6\leq \alpha\leq 1$. The opposite behavior appears in the case of hard spheres ($d=3$), although a certain scatter of the data in this case prevents us from confirming or refuting the behavior observed in the two-dimensional case.

\begin{figure}
  \includegraphics[width=\columnwidth]{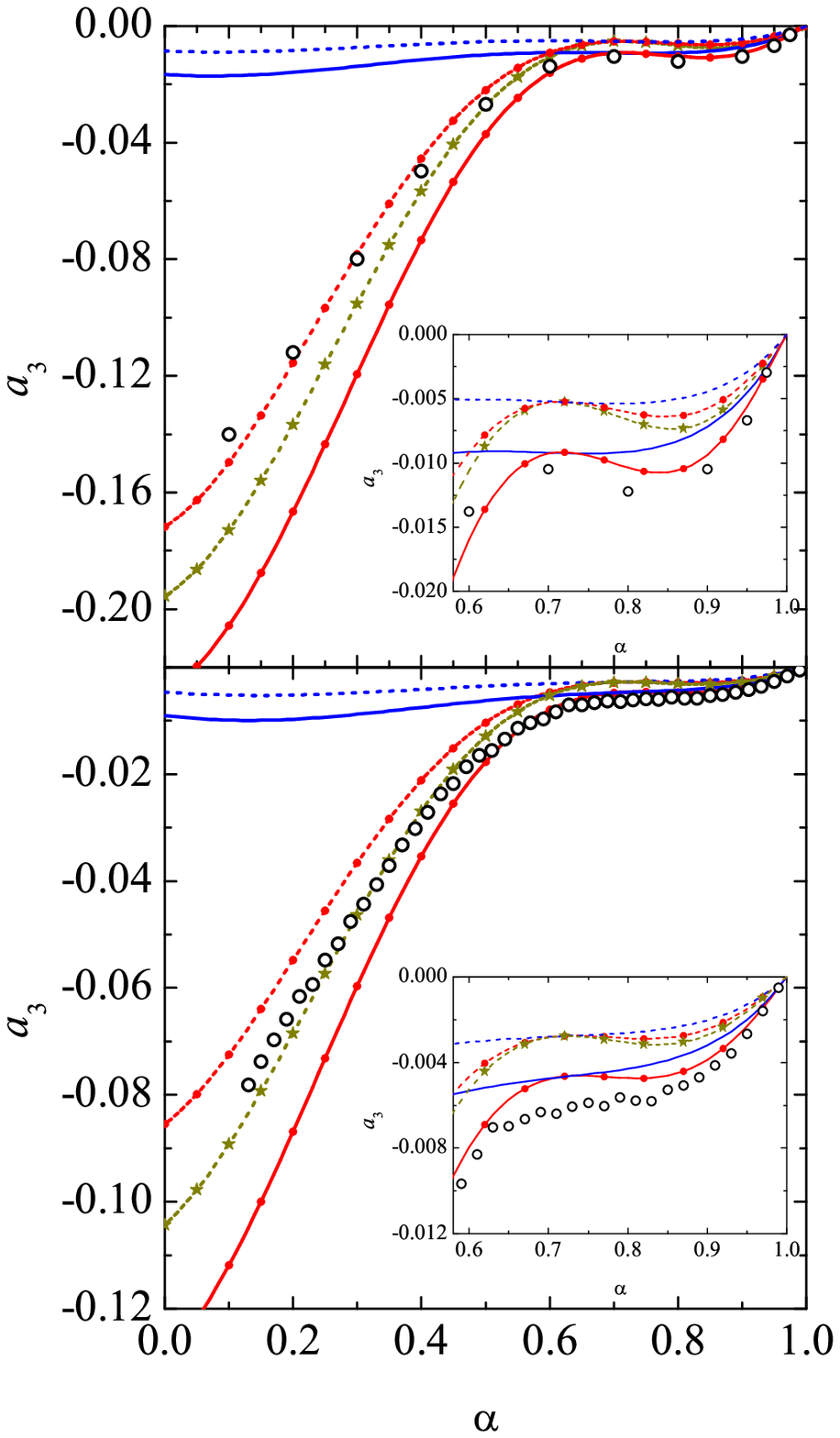}
\caption{(Color online)
Plot of the third Sonine coefficient $a_3$ as a function of the coefficient of restitution $\alpha$ for $d=2$ (top panel) and $d=3$ (bottom panel).
The circles represent DSMC results ($d=2$: this work; $d=3$:
Ref.\ \protect\cite{BP06}), while the lines correspond to $a_3^\Ia$
 (- -$\bullet$- -),
$a_3^\Ib$ (- - -), $a_3^\Ih$
(- -$\star$- -),
$a_3^\IIa$
(---$\bullet$---),
and $a_3^\IIb$
(------). The insets magnify the region $0.6\leq\alpha\leq 1$
}
\label{fig2}       
\end{figure}

Next, we consider the third Sonine coefficient $a_3$. The simulation data are compared with $a_3^\Ia$,  $a_3^\Ib$, $a_3^\Ih$, $a_3^\IIa$, and $a_3^\IIb$ in Fig.\ \ref{fig2}. It is apparent that both approaches $a_3^\Ib$ and $a_3^\IIb$ have a very poor global performance, failing to account for the rapid increase of the magnitude of $a_3$ when $\alpha\lesssim 0.6$. However, a good semi-quantitative agreement with computer simulations is found for $a_3^\Ia$, $a_3^\Ih$, and $a_3^\IIa$. All of this implies that the linearization of Eq.\ \eqref{12bis} with $p=3$ is much less accurate than the linearization of Eq.\ \eqref{12} with $p=3$, in contrast to the situation with $p=2$. Interestingly enough, among the three estimates of $a_3$ based on the linearization of Eq.\ \eqref{12} with $p=3$, the best behavior is presented by the two class-I approximations, namely $a_3^\Ia$ for $d=2$ and $a_3^\Ih$ for $d=3$.
As for the region $0.6\leq \alpha\leq 1$, the two class-II approximations are the most accurate ones. This can be understood as a consequence of the better behavior of $a_2^\IIa$ and $a_2^\IIb$ over  $a_2^\Ia$ and $a_2^\Ib$ in that region, as discussed above in connection with Fig.\ \ref{fig1}.

%

The performance of the five linear approximations is succintly summarized in Table \ref{table1}. The best global agreement with simulations is achieved by the approximation ``Ih'', i.e., $a_2$ is autonomously obtained by linearizing Eq.\ \eqref{12bis} with $p=2$ and neglecting $a_3$. Next, $a_3$ is obtained in terms of $a_2$ and $\alpha$ from the linearization of Eq.\ \eqref{12} with $p=3$.
The second best combination is ``IIa'', where $a_2$ and $a_3$ are simultaneously derived from linearization of Eq.\ \eqref{12}  with $p=2$ and $p=3$. While ``Ih'' is simpler than ``IIa'', it provides a better estimate of $a_2$ and $a_3$ for high inelasticity ($\alpha\lesssim 0.6$). This comes from a fortunate cancellation of errors and is yet another indication on the non-negligible character of nonlinear terms and higher-order Sonine coefficients in that region \cite{BP06}. On the other hand, the best estimate of $a_3$ for $0.6<\alpha<1$ is provided by $a_3^\IIa$.

\begin{table*}
\caption{DSMC values \protect\cite{MS00} of $a_2$, $\mu_2$, $\mu_4$, $\delta{\mu}_2$, $\delta\widetilde{\mu}_2$, $\delta{\mu}_4$, and $\delta\widetilde{\mu}_4$ for $d=3$}
\label{table2}       
\begin{tabular}{cccccccc}
\hline\noalign{\smallskip}
$\alpha$ &$a_2$&$\mu_2$&$\mu_4$&$\delta{\mu}_2$&$\delta\widetilde{\mu}_2$&$\delta{\mu}_4$&$\delta\widetilde{\mu}_4$\\
\noalign{\smallskip}\hline\noalign{\smallskip}
$0.8$&$-0.0141$&$0.8950$&$4.414$&$-0.005$&$-0.005$&$-0.01$&$-0.01$\\
$0.6$&$0.0207$ &$ 1.6101$&$8.213$&$0.000$&$0.000$&$0.03$&$0.02$\\
$0.4$&$ 0.0760$&$ 2.1354$&$11.494$&$0.000$&$0.002$&$0.09$&$0.02$\\
$0.2$&$0.1274$&$2.4625$&$13.881$&$-0.001$&$0.006$&$0.20$&$0.02$\\
\noalign{\smallskip}\hline
\end{tabular}
\end{table*}

Let us now define the deviations
\beq
\delta\mu_2\equiv \mu_2-\Ll{\mu_2},
\label{34}
\eeq
\beq
\delta\widetilde{\mu}_2\equiv \mu_2(1+a_2)-\Ll{\mu_2(1+a_2)},
\label{34t}
\eeq
\beq
\delta\mu_4\equiv \mu_4-\Ll{\mu_4},
\label{35}
\eeq
\beq
\delta\widetilde{\mu}_4\equiv \frac{\mu_4}{1+a_2}-\Ll{\frac{\mu_4}{1+a_2}}.
\label{35t}
\eeq
Consequently,
\beq
\Ll{\mu_4-2\mu_2\frac{\langle c^4\rangle}{\langle c^2\rangle}}=(d+2)\delta\widetilde{\mu}_2-\delta{\mu}_4,
\label{36}
\eeq
\beq
\Ll{\frac{\mu_4}{\langle c^4\rangle}-2\frac{\mu_2}{\langle c^2\rangle}}=\frac{4}{d(d+2)}\left[(d+2)\delta{\mu}_2-\delta\widetilde{\mu}_4\right].
\label{37}
\eeq
The fact that $a_2^\Ib=a_2^\Ih$ exhibits a better agreement with simulations than $a_2^\Ia$ in the high-inelasticity region obviously implies that $(d+2)\delta{\mu}_2-\delta\widetilde{\mu}_4\approx 0$ is a better approximation than $(d+2)\delta\widetilde{\mu}_2-\delta{\mu}_4\approx 0$ in that region. In principle, this does not necessarily mean that $\delta{\mu}_2\approx 0$ and $\delta\widetilde{\mu}_4\approx 0$ are better approximations than $\delta\widetilde{\mu}_2\approx 0$ and $\delta{\mu}_4\approx 0$, respectively, since a certain cancellation of terms might occur in the difference $(d+2)\delta{\mu}_2-\delta\widetilde{\mu}_4$. To clarify this point, Table \ref{table2} shows $a_2$, $\mu_2$, $\mu_4$, $\delta{\mu}_2$, $\delta\widetilde{\mu}_2$, $\delta{\mu}_4$, and $\delta\widetilde{\mu}_4$ as obtained from DSMC simulations for inelastic hard spheres ($d=3$) \cite{MS00}. We can observe that one typically has $|\delta\mu_2|<|\delta\widetilde{\mu}_2|$ and $|\delta\widetilde{\mu}_4|<|\delta{\mu}_4|$. Moreover, $|\delta{\mu}_2|$ and $|\delta\widetilde{\mu}_2|$ are considerably smaller than $|\delta{\mu}_4|$ and $|\delta\widetilde{\mu}_4|$, thus implying that the errors made when linearizing $\mu_4$ or $\mu_4/(1+a_2)$ are generally larger than those made when linearizing $\mu_2$ or $\mu_2(1+a_2)$. Therefore, the property  $|\delta\widetilde{\mu}_4|<|\delta{\mu}_4|$ is sufficient to justify that the estimate of $a_2$ obtained by setting  $\delta\mu_2\to 0$ and $\delta\widetilde{\mu}_4\to 0$ in Eq.\ \eqref{37} is more accurate than the one obtained by setting  $\delta\widetilde{\mu}_2\to 0$ and $\delta{\mu}_4\to 0$ in Eq.\ \eqref{36}.

\section{White-noise thermostat\label{sec5}}
Thus far we have assumed granular gases in the freely cooling state. The scaled quantities in this state are fully equivalent to those  of granular gases kept in a steady state by a Gaussian thermostat \cite{MS00}, i.e., by the action of a deterministic non-conservative force proportional to the particle velocity. On the other hand, a popular way of mimicking agitated granular gases is by means of a stochastic force assumed to have the form of a Gaussian white noise \cite{WM96,SE03}.

A simple estimate of $a_2$ in the case of the white-noise thermostat was derived by van Noije and Ernst \cite{vNE98} and shown to be in excellent agreement with computer simulations \cite{MS00}. However, to the best of our knowledge, an analytical expression for $a_3$ has not been proposed so far. The aim of this section is to fill this gap by applying linear approximations and exploiting the knowledge of the coefficients $A_i$, $B_i$, and $C_i$ in Eqs.\ \eqref{15}--\eqref{17} \cite{BP04,vNE98,BP06}. The starting point is the moment hierarchy, which now reads \cite{vNE98,MS00}
\beq
\mu_{2p}=p\frac{d+2p-2}{d}\mu_2 {\langle c^{2p-2}\rangle}, \quad p\geq 2,
\label{38}
\eeq
or, equivalently,
\beq
\frac{\mu_{2p}}{\langle c^{2p-2}\rangle}=p\frac{d+2p-2}{d}\mu_2 , \quad p\geq 2.
\label{39}
\eeq
In the class-I approximation, one takes $p=2$ and linearizes with respect to $a_2$, i.e.,
\beq
\Ll{\mu_4-(d+2)\mu_2}=0.
\label{40}
\eeq
This condition is independent of whether we linearize Eq.\ \eqref{38} or Eq.\ \eqref{39}, in contrast to the free cooling case. The solution of Eq.\ \eqref{40} is
\beqa
a_2^\Ia=a_2^\Ib&=&\frac{B_0-(d+2)A_0}{(d+2)A_2-B_2}\nn
&=&\frac{16(1-\alpha)(1-2\alpha^2)}{73+56d-3 (35+8
d)\alpha+30(1-\alpha)\alpha^2}.
\label{41}
\eeqa
This is the result obtained in Ref.\ \cite{vNE98}. Once $a_2$ is known, $a_3$ is determined from
\beq
\LL{\mu_6-3\frac{d+4}{d}\mu_2\langle c^4\rangle}=0
\label{42}
\eeq
in the approximation ``Ia'' or from
\beq
\LL{\frac{\mu_6}{\langle c^4\rangle}-3\frac{d+4}{d}\mu_2}=0
\label{43}
\eeq
in the approximation ``Ib''. The results are
\beq
a_3^\Ia(\alpha)=G_a(\alpha,a_2^\Ia(\alpha)),
\label{n8wn}
\eeq
\beq
a_3^\Ib(\alpha)=G_b(\alpha,a_2^\Ia(\alpha)),
\label{n9wn}
\eeq
where
\beqa
G_a(\alpha,a_2)&\equiv&
\Big\{C_0-\frac{3}{4}(d+2)(d+4)A_0+\Big[C_2-\frac{3}{4}(d+2)\nn
&&\times(d+4)(A_0+A_2)\Big]a_2\Big\}/\Big[\frac{3}{4}(d+2)(d+4)\nn
&&\times A_3-C_3\Big],
\label{n10wn}
\eeqa
\beqa
G_b(\alpha,a_2)&\equiv&
\Big\{C_0-\frac{3}{4}(d+2)(d+4)A_0+\Big[C_2-C_0\nn
&&-\frac{3}{4}(d+2)
(d+4)A_2\Big]a_2\Big\}/\Big[\frac{3}{4}(d+2)\nn
&&\times(d+4)A_3-C_3\Big].
\label{n11wn}
\eeqa
\begin{figure}
  \includegraphics[width=\columnwidth]{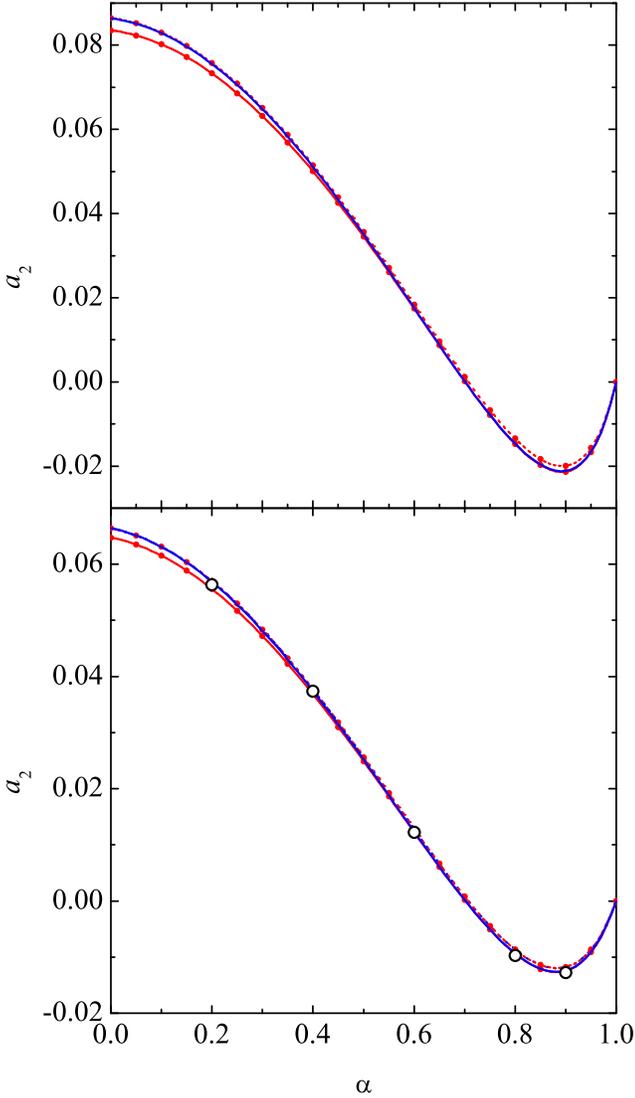}
\caption{(Color online)
Plot of the second Sonine coefficient $a_2$ as a function of the coefficient of restitution $\alpha$ in the case of the white-noise thermostat for $d=2$ (top panel) and $d=3$ (bottom panel).
The circles in the bottom panel represent DSMC results from
Ref.\ \protect\cite{MS00}, while the  lines correspond to $a_2^\Ia=a_2^\Ib$
 (- -$\bullet$- -),
$a_2^\IIa$
(---$\bullet$---),
and $a_2^\IIb$
(------).
}
\label{fig3}       
\end{figure}
%
\begin{figure}
  \includegraphics[width=\columnwidth]{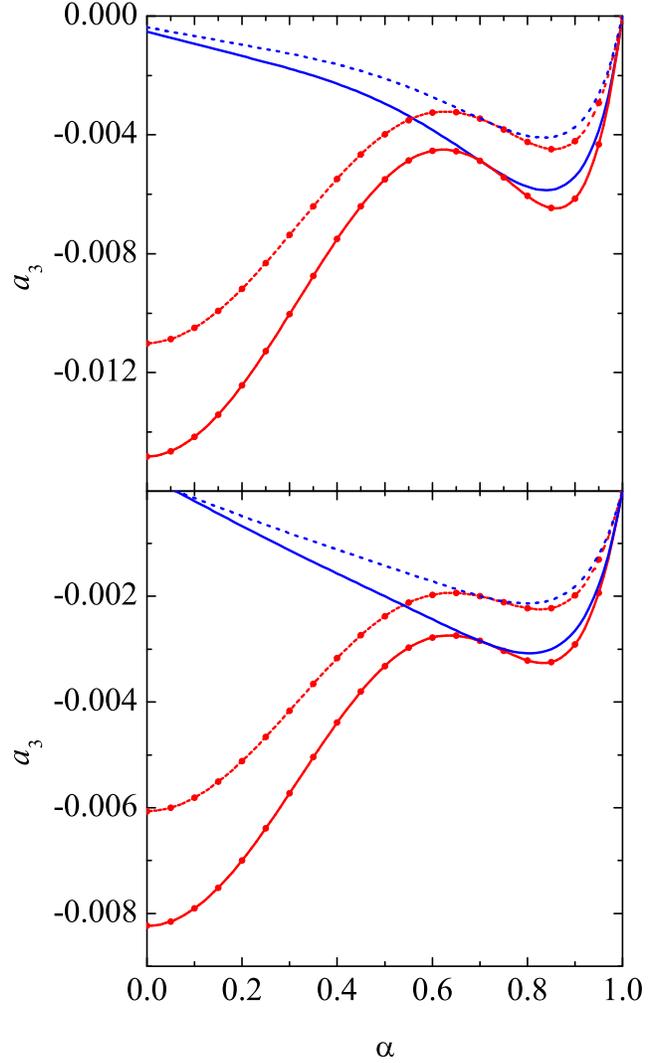}
\caption{(Color online)
Plot of the third Sonine coefficient $a_3$ as a function of the coefficient of restitution $\alpha$ in the case of the white-noise thermostat for $d=2$ (top panel) and $d=3$ (bottom panel).
The lines correspond to $a_3^\Ia$
 (- -$\bullet$- -),
$a_3^\Ib$ (- - -),
$a_3^\IIa$
(---$\bullet$---),
and $a_3^\IIb$
(------).
}
\label{fig4}       
\end{figure}

In the class-II approximations both $a_2$ and $a_3$ are simultaneously obtained from
\beq
\LL{\mu_4-(d+2)\mu_2}=0
\label{44}
\eeq
and either Eq.\ \eqref{42} (approximation ``IIa'') or Eq.\ \eqref{43} (approximation ``IIb''). The solutions are
\beq
a_2^\IIa(\alpha)=\frac{{N}_a(\alpha)}{{D}_a(\alpha)},\quad
a_3^\IIa(\alpha)=G_a(\alpha, a_2^\IIa(\alpha)),
\label{27wn}
\eeq
\beq
a_2^\IIb(\alpha)=\frac{{N}_b(\alpha)}{{D}_b(\alpha)},\quad
a_3^\IIb(\alpha)=G_b(\alpha, a_2^\IIb(\alpha)),
\label{33wn}
\eeq
where
\beqa
{N}_a(\alpha)&\equiv& {C_3}{B_0}- {C_0}{B_3} +
    (d+2)( A_3 C_0- A_0 C_3)  +
       \frac{3}{4}(d+2)\nn
       &&\times
       (d+4)(A_0 B_3-  {A_3}{B_0}),
       \label{28wn}
\eeqa
\beqa
     {D}_a(\alpha)&\equiv& C_2 B_3-C_3 B_2 +
   (d+2)( {A_2}C_3-A_3 C_2)\nn
   &&+\frac{3}{4}(d+2)(d+4)
       [A_3B_2-A_2B_3\nn
       &&+(d+2)A_0A_3-A_0B_3],
       \label{29wn}
\eeqa
\beq
{N}_b(\alpha)={N}_a(\alpha),
       \label{31wn}
\eeq
\beqa
    {D}_b(\alpha)&\equiv& (C_2-C_0) B_3-C_3 B_2 +
   (d+2)[ {A_2}C_3-A_3 (C_2\nn
   &&-C_0)]+\frac{3}{4}(d+2)(d+4)
       (A_3B_2-A_2B_3).
       \label{32wn}
\eeqa

Figure \ref{fig3} shows the $\alpha$-dependence of $a_2^\Ia=a_2^\Ib$, $a_2^\IIa$, and $a_2^\IIb$. The three curves are very close each other, which indicates that $a_2^2$ and $a_k$ with $k\geq 3$ are indeed small enough to be neglected in $\mu_4=(d+2)\mu_2$. It is interesting to note that $a_2^\IIa$ and $a_2^\IIb$ are practically identical in the region $0.6\leq \alpha\leq 1$, where they are slightly more accurate (at least in the three-dimensional case) than $a_2^\Ia=a_2^\Ib$. On the other hand, $a_2^\Ia=a_2^\Ib$ and $a_2^\IIb$ are practically indistinguishable in the region of small $\alpha$. The theoretical predictions $a_3^\Ia$, $a_3^\Ib$, $a_3^\IIa$, and $a_3^\IIb$ are displayed in Fig.\ \ref{fig4}. Although there are no simulation data to compare with, it seems plausible to conjecture that the trends observed in Fig.\ \ref{fig2} are repeated now: the two class-II approximations are more accurate than the two class-I approximations for small dissipation, $a_3^\IIa$ being possibly better than $a_3^\IIb$, while the two ``b'' approximations are rather poor at high dissipation. Note that, since $a_2^\Ia=a_2^\Ib$ in the case of the white-noise thermostat, the hybrid approximation ``Ih'' coincides with ``Ia'', i.e., $a_3^\Ih=a_3^\Ia$. A feature that becomes apparent when comparing Figs.\ \ref{fig1} and \ref{fig2} with Figs.\ \ref{fig3} and \ref{fig4}, respectively, is that the magnitudes of $a_2$ and $a_3$ in the white-noise case are about twice and ten times, respectively, smaller than those in the freely cooling state. This is closely related to the fact that the overpopulation of the high-energy tail is much smaller in the former case than in the latter. More specifically, instead of Eq.\ \eqref{tail}, now we have \cite{vNE98}
\beq
F(\mathbf{c})\sim e^{-\frac{2}{3}\sqrt{2\xi c^3}},
\label{tail_wn}
\eeq
where $\xi$ is the same quantity as in Eq.\ \eqref{tail}.

\section{Conclusions\label{sec6}}

\begin{figure}
  \includegraphics[width=\columnwidth]{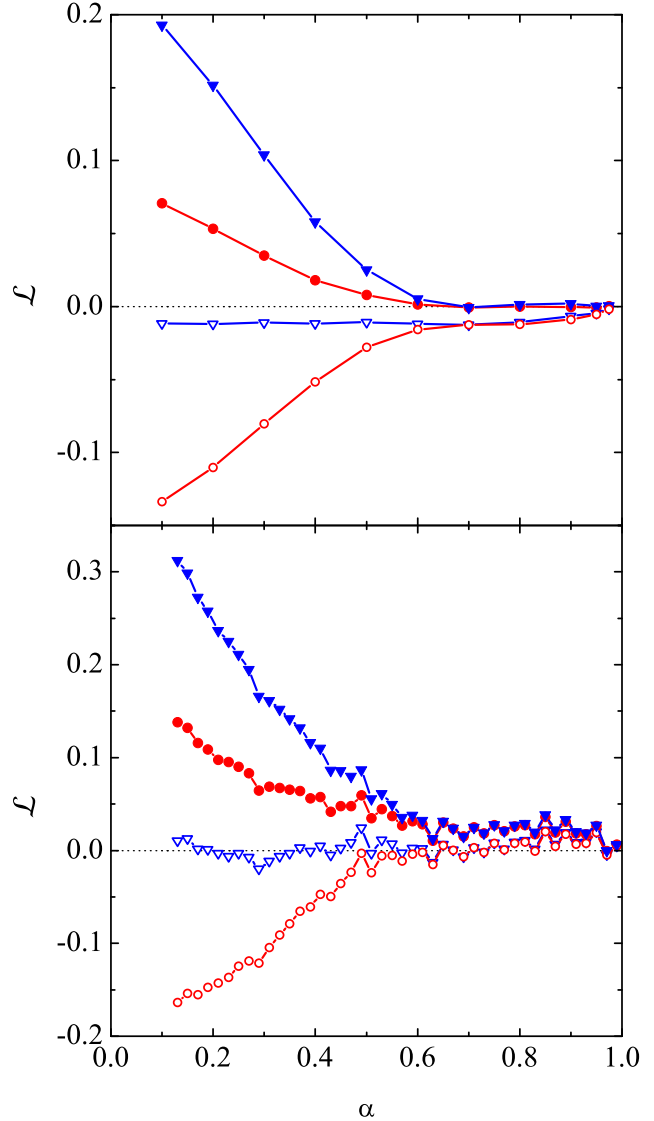}
\caption{(Color online)
Plot of $\Ll{\mu_4-(d+2)\mu_2(1+a_2)}$ ($\circ$), $\LL{\mu_4-(d+2)\mu_2(1+a_2)}$ ($\bullet$), $\Ll{\mu_4/(1+a_2)-(d+2)\mu_2}$ ($\triangledown$), and $\LL{\mu_4/(1+a_2)-(d+2)\mu_2}$ ($\blacktriangledown$) for $d=2$ (top panel) and $d=3$ (bottom panel). The symbols are obtained from  DSMC results of $a_2$ and $a_3$ ($d=2$: this work; $d=3$:
Ref.\ \protect\cite{BP06}). The lines are guides to the eye.
}
\label{fig5}       
\end{figure}

{The second and third  coefficients in the Sonine polynomial expansion of the (scaled) velocity distribution function $F(\mathbf{c})$ of a granular gas  characterize the deviation of $F(\mathbf{c})$ from the Maxwellian and are important, for instance, in the precise determination of transport coefficients. While for practical purposes an interval $0.8\lesssim \alpha<1$ for the coefficient of normal restitution is sufficient, it becomes necessary from a more fundamental viewpoint to consider the whole range $0<\alpha<1$.}

The Sonine coefficients $a_2(\alpha)$ and $a_3(\alpha)$ can be measured in computer simulations (e.g., DSMC), so one could in principle make a least-square fit to  certain functional forms. However, this would not be satisfactory from a fundamental point of view and would not provide any insight into  the intricacies of $F(\mathbf{c})$ and its Sonine expansion. It is then much more challenging to devise theoretical approximations that can be subsequently assessed by comparison with simulation results.

{In this paper we have been mainly concerned with a family of linear approximations to estimate $a_2(\alpha)$ and $a_3(\alpha)$ in the freely cooling state. We have found that a good compromise} between accuracy and simplicity is represented by the hybrid approximation here denoted as ``Ih'', in which the second and third Sonine coefficients are given by Eqs.\ \eqref{24} and \eqref{n12}, respectively. {For the benefit of the reader here we quote the final and complete expressions:}
\beq
a_2(\alpha)=
\frac{16(1-\alpha)(1-2\alpha^2)}{25+24d- (57-
8d)\alpha-2(1-\alpha)\alpha^2},
\label{F1}
\eeq
\beq
a_3(\alpha)=-\frac{16 a_2(\alpha)}{1-2\alpha^2}\frac{P_\text{HCS}(\alpha)}{Q_\text{HCS}(\alpha)},
\label{F2}
\eeq
{where}
\beqa
P_\text{HCS}(\alpha)&=&167 + 50 d - (191 + 26 d) \alpha - 2 (307 + 100 d) \alpha^2 \nn
&&+
 2 (339 + 68 d) \alpha^3 + 32 (16 + 7 d) \alpha^4\nn&& -
 32 (18 + 5 d) \alpha^5 + 144 (1- \alpha)\alpha^6,
 \label{F3}
 \eeqa
\beqa
Q_\text{HCS}(\alpha)&=&521 +
 1396 d + 368 d^2 - (1481 + 820 d  \nn&& - 16 d^2)\alpha+
 4 (583 + 262 d) \alpha^2 \nn&& - 20 (155 + 14 d) \alpha^3+
 280 (1-\alpha)\alpha^4
 \label{F4}
\eeqa
However, if more precise values in the domain $0.6\leq \alpha<1$ are really needed, it might be preferable to consider  the more complicated approximation ``IIa'' given by Eqs.\ \eqref{26} and \eqref{27} \cite{BP06}.

On the other hand, It is known that  in the high-inelasticity region $\alpha\lesssim 0.6$ (i.e., once $a_2$ becomes positive) the higher-order Sonine coefficients are no longer negligible \cite{NBSG07,HOB00,BP06}, so that the linear approximations based on the neglect of nonlinear terms and of $a_k$ with $k\geq 3$ or $k\geq 4$ are not \emph{a priori} reliable. This is made evident by the lack of self-consistency of different linear approximations used to estimate $a_2$ from the first non-trivial equation of the moment hierarchy, as shown in Fig.\ \ref{fig1}. What is indeed surprising is that the simple linear approximation \eqref{F1} provides such an excellent estimate both for $d=2$ and $d=3$. This means that, even though $a_2^2$, $a_3$, $a_4$, \ldots are not negligible at all if $\alpha\lesssim 0.6$, somehow they  practically cancel out in the combination $\mu_4/\langle c^4\rangle-2 \mu_2/\langle c^2\rangle$, while  still playing a significant role in the combination $\mu_4-2 \mu_2\langle c^4\rangle/\langle c^2\rangle$. This is clearly seen in Fig.\ \ref{fig5}, where we plot $\Ll{\mu_4-(d+2)\mu_2(1+a_2)}$, $\LL{\mu_4-(d+2)\mu_2(1+a_2)}$, $\Ll{\mu_4/(1+a_2)-(d+2)\mu_2}$, and $\LL{\mu_4/(1+a_2)-(d+2)\mu_2}$ by using the simulation data of $a_2$ and $a_3$. While the magnitude of the other three quantities rapidly increases with increasing inelasticity if $\alpha\lesssim 0.6$, that of $\Ll{\mu_4/(1+a_2)-(d+2)\mu_2}$ remains small and practically constant. This property might be useful to contribute to a better understanding of the full velocity distribution function $F(\mathbf{c})$.

{Although this paper has focused on the homogeneous cooling state, the analysis has been straightforwardly extended in Sec.\ \ref{sec5} to  a granular gas heated by a white-noise thermostat. In that case, the optimal combination of estimates is provided by Eqs.\ \eqref{41} and \eqref{n8wn}. More specifically,}
\beq
a_2=\frac{16(1-\alpha)(1-2\alpha^2)}{73+56d-3 (35+8
d)\alpha+30(1-\alpha)\alpha^2},
\label{F5}
\eeq
\beq
a_3(\alpha)=-\frac{16 a_2(\alpha)}{1-2\alpha^2}\frac{P_\text{WN}(\alpha)}{Q_\text{WN}(\alpha)},
\label{F6}
\eeq
{where}
\beqa
P_\text{WN}(\alpha)&=&67 + 10 d - 7 (13 - 2 d) \alpha - 2 (119 + 20 d) \alpha^2 \nn
&&+
 2 (151 - 12 d) \alpha^3 + 32 (8 + 3 d) \alpha^4\nn&& -
 32 (10 + d) \alpha^5 + 80 (1 - \alpha) \alpha^6,
\label{F7}
\eeqa
\beqa
Q_\text{WN}(\alpha)&=&2569 + 2932 d + 624 d^2 - (3529 + 2356 d \nn&&
+ 240 d^2 ) \alpha +
 4 (583 + 262 d) \alpha^2 \nn&&- 20 (155 + 14 d) \alpha^3 +
 280 (1 - \alpha) \alpha^4.
 \label{F8}
\eeqa
It would be interesting to test the accuracy of Eq.\ \eqref{F6} against computer simulations. 

\appendix
\section{Expressions for $A_i$, $B_i$, and $C_i$\label{appA}}
The explicit expressions of the coefficients $A_i$, $B_i$, and $C_i$
as functions of $d$ and $\alpha$ are \cite{BP04,vNE98,BP06}
\beq
A_0=K(1-\alpha^2),\quad A_2=\frac{3K}{16}(1-\alpha^2),\quad
A_3=\frac{K}{64}(1-\alpha^2),
\label{18}
\eeq
\beq
B_0=K(1 - \alpha^2) \left(d + \frac{3}{2} + \alpha^2\right),
\eeq
\beq
B_2=K(1+\alpha)\left[\frac{3}{32} (1 - \alpha) (10 d + 39 + 10 \alpha^2) + (d -
    1)\right],
    \eeq
\beq
B_3=-\frac{K}{128}(1+\alpha)\left[(1 - \alpha) (97 + 10 \alpha^2) +2(d -
1)(21 - 5 \alpha)\right],
\eeq
\beq
C_0=
 \frac{3K}{4} (1 - \alpha^2) \left[(d + \alpha^2) (5 + 2 \alpha^2) + d^2 +
    \frac{19}{4}\right],
    \eeq
    \beq
C_2=
 \frac{3K}{256} (1 - \alpha^2) \left[1289 +
     4 (d + \alpha^2) (311 + 70 \alpha^2) + 172 d^2\right] +
  \frac{3}{4} {\beta},
  \eeq
  \beq
C_3= -\frac{3K}{1024} (1 - \alpha^2) \left[2537 +
     4 (d + \alpha^2) (583 + 70 \alpha^2) + 236 d^2\right] -
  \frac{9}{16} \beta,
  \eeq
where
\beq
K\equiv \frac{\pi^{(d-1)/2}}{\sqrt{2}\Gamma(d/2)},\quad
\beta\equiv K(1 + \alpha) \left[(d - \alpha) (3 +
       4 \alpha^2) + 2 (d^2 - \alpha)\right].
       \eeq

\section{Other class-I linear approximations for $a_2$\label{appB}}
As a generalization of Eqs.\ \eqref{n1I} and \eqref{n1bI}, let us consider the family of approximations
\beq
\Ll{\frac{\mu_4^{1-z}}{\langle
c^4\rangle^x\mu_2^y}-2\frac{\mu_2^{1-y}\langle
c^4\rangle^{1-x}}{\langle
c^2\rangle\mu_4^z}}=0
\label{B1}
\eeq
and let us denote by $a_2^{(x,y,z)}$ the associated solution. In particular, $a_2^{(0,0,0)}\equiv
a_2^\Ia$ and $a_2^{(1,0,0)}\equiv a_2^\Ib$. The
eight possibilities considered by Coppex et al. \cite{CDPT03} correspond to
$x=0,1$, $y=0,1$, and $z=0,1$.  It is  easy to
check that the solution to Eq.\ \eqref{B1} is
\beq
a_2^{(x,y,z)}=\frac{a_2^\Ia}{1+h_{xyz}a_2^\Ia},
\label{B2}
\eeq
where
\beq
h_{xyz}\equiv x+y\frac{A_2}{A_0}+z\frac{B_2}{B_0}.
\label{B3}
\eeq
Equation \eqref{B2} is a generalization of Eq.\ \eqref{n7}.

Of course, other alternative possibilities exist. For instance, one
can generalize Eq.\ \eqref{B1} to
\beq
\Ll{\Phi\left(\frac{\mu_4^{1-z}}{\langle
c^4\rangle^x\mu_2^y}\right)-\Phi\left(2\frac{\mu_2^{1-y}\langle
c^4\rangle^{1-x}}{\langle
c^2\rangle\mu_4^z}\right)}=0,
\label{B4}
\eeq
where $\Phi(X)$ is an arbitrary function. The corresponding approximation for $a_2$ will depend on the choice of $\Phi(X)$, apart from the values of $x,y,z$.

\begin{acknowledgements}
The authors are grateful  to N. Brilliantov and T. P\"oschel for providing the simulation data of Ref.\ \cite{BP06}.
This research  has been supported by the Ministerio de Educaci\'on y
Ciencia (Spain) through grants Nos.\ FIS2007-60977 (A.S.) and  DPI2007-63559 (J.M.M.), partially financed by FEDER funds.
\end{acknowledgements}



\end{document}